\documentclass[aps,prl,reprint,twocolumn,superscriptaddress,showpacs,floatfix,longbibliography]{revtex4-2}
\usepackage[latin1]{inputenc}
\usepackage{epsfig}
\usepackage{amsfonts}
\usepackage{amssymb}
\usepackage{calrsfs}
\usepackage{amsmath}
\usepackage{xcolor}
\usepackage{graphicx}
\usepackage[colorlinks=true,allcolors=blue]{hyperref}
\hypersetup{breaklinks=true}
\usepackage{lipsum}
\usepackage{bm}
\usepackage[mathscr]{euscript}

\begin{document}
	
	\title{Periodically driven thermodynamic systems under vanishingly small viscous drives}
	\author{Shakul Awasthi}
	\email{shakul23@kias.re.kr}
	\affiliation{School of Physics, Korea Institute for Advanced Study, Seoul 02455, Korea}
	\author{Sreedhar B. Dutta}
	\email{sbdutta@iisertvm.ac.in}
	\affiliation{School of Physics, Indian Institute of Science Education and Research Thiruvananthapuram, Thiruvananthapuram 695551, India}
	
	
	\begin{abstract} 
		\noindent 
		Periodically driven thermodynamic systems support stable non-equilibrium oscillating states with properties drastically different from equilibrium. They exhibit even more exotic features for low viscous drives, which is a regime that is hard to probe due to singular behavior of the underlying Langevin dynamics near vanishing viscosity. We propose a method, based on singular perturbation and Floquet theories, that allows us to obtain oscillating states in this limit. We then find two distinct classes of distributions, each exhibiting interesting features that can be exploited for a range of practical applicability, including cooling a system and triggering chemical reactions through weakly interacting driven environments.

	\end{abstract}

	\maketitle
	
	\newcommand{\bee}{\begin{equation}}
		\newcommand{\eee}{\end{equation}}
	\newcommand{\tm}{(t)}
	\newcommand{\gm}{\gamma}
	\newcommand{\dbar}{d\hspace*{-0.08em}\bar{}\hspace*{0.1em}}
	\newcommand{\haf}{\frac{1}{2}}
	\newcommand{\x}{\mathbf{x}}
	
	%
	%
	
	\textsl{Introduction.--} Langevin framework provides a simple and suitable description for periodically driven thermodynamic systems and allows us to effectively probe their properties beyond equilibrium~\cite{Kampen1992,Risken1996,Zinn2021}.
	It is therefore not surprising to witness the current proliferation of studies on periodically driven Langevin systems knowing their relevance not only for thermodynamic systems like heat engines~\cite{Quinto2014, Martinez2017, Pietzonka2018, Lu2022} and nano-mechanical resonators~\cite{Cleland2002, Unterreithmeier2010, Eom2011} but also for wider range of applications including climate modeling~\cite{Palmer2019, Margazoglou2021, Held1982, Wiesenfeld1995, Lucarini2019}, biological processes~\cite{Schmiedl2007, Jain2022} and ecological trends~\cite{Kuehn2013, Rozenfeld2001, Ma2023}.
	
	In this Letter, we focus our study, for multiple reasons, on a periodically driven under-damped Brownian particle moving in small viscous regime. Firstly, Brownian particle is a paradigmatic Langevin system that can aid in extracting general nonequilibrium features of macroscopic systems~\cite{Risken1996,Kampen1992}. Secondly, inertial effects can become significant with increasing drive frequency.
	Thirdly, driven Brownian particle is known to exist in stable nonequilibrium states including oscillating states~\cite{Jung1993,Koyuk2019, Proesmans2016, Awasthi2020, Awasthi2021}. Fourthly, the limit of vanishing viscosity is singular and of immense interest both for mathematical and physical reasons, including in the study of inviscid flows~\cite{Ludford1960, Friedlander1991, Prakash2013} and in understanding the eluding turbulence~\cite{Sreenivasan1999,Campbell1985}.  
		
	 Periodically driven Langevin systems do not relax to equilibrium, and instead can support non-equilibrium states, referred to as oscillating states (OS)~\cite{Koyuk2019,Proesmans2016, Awasthi2020,Awasthi2021,Awasthi2022,Chen2023}. Unlike in equilibrium, for a given bath temperature OS are not independent of viscosity and can carry significant viscous memory and in turn exhibit drastically distinct thermodynamic properties. Though the effects of viscosity in various stochastic systems are extensively studied over the whole gamut~\cite{Huang2005,Stevens2014}, there is in general a bias towards large viscous or over-damped regimes~\cite{Hiss1973,Sweat2017,Ding2002,Okuyama1986}. The systems typically relax fast in this regime and are easy to control, numerically or perturbatively, their convergence to the asymptotic state. Furthermore their dynamics usually reduce to mathematically well-studied continuous Markov processes. But the physics exhibited by Langevin systems when viscosity is small~\cite{Case2008,Csernai2006,Prakash2013,Cruz2002} can be antithetical to that when it is large, particularly in the presence of drives where inertial effects do not decouple, and thus obligates dedicated study. This motivates us to address and answer the questions: How do we identify an oscillating state (OS) when the approach to asymptotic state becomes increasingly sluggish as viscosity reduces? What are the properties of OS, if they exist, in this singular limit of vanishing viscosity? Is it possible to develop a systematic perturbative scheme about the singular limit? How distinct are low viscous OS? What novel mechanisms and applications can their study lead us to?

	%
	%
	\textsl{Model.--}We consider a Brownian particle in harmonic potential described by an under-damped Langevin equation  
	\begin{align}\label{stoc-dyn}
		\dot{X}_t = V_t ~,\quad
		\dot{V}_t = -\gamma (t) V_t - k_0 X_t + \eta(t)~,
	\end{align} 
	where $- k_0 X_t$ denotes the external force parameterized by$~k_0$. The particle experiences$~-\gamma(t)V_t$ viscous drag and a random Gaussian noise$~\eta(t)$ with zero mean and $\langle \eta_t \eta_{t'} \rangle = 2 D(t) \delta(t-t')$. We choose$~k_0$ to be constant while$~\gamma(t)$ and$~D(t)$ are time-dependent functions with period$~T$.
	
	In order to investigate the influence of drives on asymptotic states in low viscous regime, we extend the chosen periodic functions $\gamma(t)\to\gamma_\alpha(t)=\alpha\gamma(t)$ and $D(t)\to D_\alpha(t)=\alpha D(t)$ by a one-parameter extension labeled by a real positive number$~\alpha$. This extension keeps the bath temperature $T_b(t)~=~D(t)/\gamma(t)$ independent of$~\alpha$, and has the advantage of isolating the dependence of these non-equilibrium states on viscous drives$~\gamma_{\alpha}(t)$ for given$~T_b(t)$. We aim to first find the asymptotic states in the limit$~\alpha \to 0$, and then explore their properties in the neighborhood of$~\alpha=0$.

	The  distribution $P(x,v,t)$ associated with the random process in Eq.~\eqref{stoc-dyn}, satisfies the Fokker-Planck (FP) equation given by
	\begin{equation}\label{FP-eqn}
		\frac{\partial}{\partial t}P(x,v,t) =\mathcal{L}(x,v;g(t)) P(x,v,t)~,
	\end{equation}
	where~$g$ denotes all the parameters, including~$\gamma$ and~$D$, and the FP operator~$\mathcal{L}$ is defined as
	\begin{equation}\label{FP-op}
		\mathcal{L}(x,v;g) :=  -\frac{\partial}{\partial x}v - \frac{\partial}{\partial v}\left[ -\gm_\alpha  v - k_0 x \right]  +  D_\alpha  \frac{\partial^{2}}{\partial v^{2}} ~.
	\end{equation}
	Under certain conditions~\cite{Awasthi2020,Awasthi2021}, $P(x,v,t)$ at large times takes a time periodic form which is independent of the initial conditions. These distributions of asymptotic states, also referred to as OS, are denoted by $P_{os}(x,v,t)$ and defined as
	\begin{equation}\label{os-lim}
		P_\text{os}(x,v,t) := \lim_{N \to \infty} P(x,v,N T + t)~.
	\end{equation}

	The choice of harmonic potential$~k_0 x^2/2$ is not a restriction. The method we propose to find OS and the general results that follow thereafter are equally applicable to any (driven) polynomial potentials~$U(x;\lambda(t))=\sum_n \lambda_n(t) x^n$.

	%
	%
	
	\textsl{Method.--}The OS distribution for given $\alpha$ has a perturbative expansion
	\begin{equation}\label{os-expansion}
		P_\text{os}(t;\alpha) = P_\text{os}^{(0)}(t) + \alpha P_\text{os}^{(1)}(t) + \alpha^2 P_\text{os}^{(2)}(t) + \cdots~,
	\end{equation}
	where $x,v$ dependence is not shown explicitly for convenience. The standard procedure of obtaining perturbative solution $P(t;\alpha) = \sum_{n=0}^{\infty}\alpha^n P^{(n)}(t)$ amounts to solving the hierarchy of equations
	\begin{equation}\label{pert-hier}
		\frac{\partial}{\partial t} P^{(n)}(t) =\mathcal{L}_0 P^{(n)}(t) +\mathcal{L}_1 P^{(n-1)}(t) ~,
	\end{equation}
	where $P^{(-1)}(t)=0$, the Liouville operator$~\mathcal{L}_0$ and perturbative operator$~\mathcal{L}_1$ are the $O(1)$ and $O(\alpha)$ parts of the FP operator, respectively. The Liouville equation is a singular limit of FP equation and its solution, which has no large-time limit, depends on the initial condition. Hence the hierarchy of solutions of Eqs.~\eqref{pert-hier} depend on their initial conditions$~P_{in}^{(m)}$ and result in the full solution
	\begin{equation}\label{pert-soln}
		P(t;\alpha) = \sum_{n=0}^{\infty}\alpha^n P^{(n)}(t; \{P_{in}^{(m)}\}_{m\le n})~,
	\end{equation}
	which asymptotically will not coincide with$~P_\text{os}(t;\alpha)$. This is expected since the limits $\alpha \to 0$ and $t \to \infty$ do not commute in this singular perturbation problem. Instead we employ Floquet theory to obtain the OS distribution for finite$~\alpha$ wherein we impose periodicity on$~P(t;\alpha)$ and in turn determine the hierarchy of initial conditions$~\{P_{in}^{(m)}\}$ uniquely. This uniqueness is ensured since OS is unique, periodic and independent of the full initial condition. We need to impose periodicity to$~O(\alpha^{n+1})$ in order to determine$~\{P_{in}^{(m)}\}$ to$~O(\alpha^{n})$. 
	
	The distribution can also be obtained from its moments which satisfy ordinary differential equations. This does not mean the hurdles of singular limit can be avoided, but can be handled following the proposed method as detailed in the supplemental materials (SM)~\cite{supp}.

	%
	%
	
	\textsl{Results.--}The OS of harmonic Brownian particle is Gaussian~\cite{Awasthi2020} with zero mean and time-periodic covariance matrix$~\Sigma$ whose elements are second moments $\Sigma_{11}=\left\langle x^{2}\right\rangle$, $\Sigma_{12}=\Sigma_{21}=\langle x v\rangle$ and $\Sigma_{22}=\left\langle v^{2}\right\rangle$. 
	
	We have established that the condition $k_0>0$ is sufficient to ensure the existence of OS in the small $\alpha$ regime for any bounded positive real periodic functions $\gamma(t)$ and $D(t)$ of time-period $T=2\pi/\omega$. In fact, we can specify an implicit inequality given$~\gamma(t)$ and$~k_0$ (or$~k(t)$, a time-dependent harmonic drive) that is necessary to satisfy for the existence of OS, as specified in SM~\cite{supp}.

	The existence of OS implies that the proposed perturbative method can be employed to obtain its distribution. We find two distinct classes of distributions having differing statistical behavior at the leading order which is a consequence of whether the time scales associated with potential and drive are in tune or not. They are distinguished by whether at least one of the drives$~\gamma$ or $D$ contains any harmonic $n_0 \omega = \pm 2\sqrt{k_0}$ or not.

	\textsl{Case I}: When the potential strength$~k_0 \neq n_0^2 \omega^2/4$, for all integers$~n_0$, then we find the leading-order moments of OS to be
	\begin{align}
		\Sigma_{11}^{(0)} = \frac{1}{k_0} \frac{\overline{D}}{ \overline{\gamma}}~,\quad \Sigma_{12}^{(0)} = 0~,\quad \Sigma_{22}^{(0)} =  \frac{\overline{D}}{ \overline{\gamma}}~,
	\end{align}
	where the overline denotes time average over a period. Thus OS distributions in the limit $\alpha\to0$ are time-independent and depend only on time-averaged viscous and noise parameters. It does not mean that viscous drive has no effect for given time-dependent bath temperature since $\overline{D}= \overline{\gamma T_b}$. The time dependence and periodicity of OS are seen at the next-order. The explicit expressions of moments to first-order and their numerical verification are given in SM~\cite{supp}.
	
	We see that in general the kinetic temperature $T_s(t):= \Sigma_{22}(t)$ of the particle in OS differs from bath-temperature $T_b(t)$ and its correlation function$~\Sigma_{12}(t)$ is non-zero. Though these observables distinguish OS from equilibrium, there are other non-equilibrium variables, such as house-keeping heat flux, entropy flux and entropy production, whose relevance is evident when we view OS as a cyclic process governed by stochastic thermodynamics.  

	In an infinitesimal time $dt$ of the cyclic process, at the level of an individual Langevin trajectory the change in energy $d E_t = \dbar Q + \dbar W$~\cite{Sekimoto2010}. The work done on the particle $\dbar W := \dot{g}\cdot \partial_g E_t dt$ depends on periodic drives$~g(t)$ and thus vanishes in absence of potential drive. The heat received from the bath is $\dbar Q_t :=  - \gamma V_t^2 dt + V_t \circ dB_t$,
	where ${dB_{t} =\int_t^{t+dt}dt' \eta(t') }$ is the Brownian noise and $\circ$ denotes Stratonovich product.  We find the rate of heat dissipation
		\begin{align}\label{hk-heat}
		q_\text{hk} = -\frac{\langle \dbar Q_t \rangle}{dt}&= \gamma_\alpha \langle V_t^2 \rangle - D_\alpha
		=\alpha \gamma\left( T_s - T_b  \right),
	\end{align}
	linearly depends on instantaneous difference of the two temperatures. This dissipation is indispensable to maintain the particle in given OS and hence is referred to as house-keeping heat flux. The Langevin dynamics also determines the rate of change of entropy 
	\begin{eqnarray}\label{s-rate}
		\frac{dS}{dt}=\frac{\gamma_\alpha}{D_\alpha} \frac{\langle \dbar Q_t \rangle}{dt} +  \frac{1}{D_\alpha} \left\langle \left( \frac{\hat{J}_v^\text{ir}}{\hat{P}_\text{os}} \right)^2_t \right\rangle ~, 
	\end{eqnarray}
	where \hbox{$S$ ~$= \langle - \log \hat{P}_{os}\rangle$}~\cite{Seifert2005} and the stochastic variables $\hat{P}_{os} = P_{os} (X_t,V_t,t)$ and $\hat{J}_v^\text{ir}= J_v^\text{ir}(X_t,V_t,t)$ are associated with $P_{os}(x,v,t)$ and the irreversible part of probability current $
		J_v^\text{ir} (x,v, t) = - \left( \gamma_\alpha v  + D_\alpha \partial_v\right) P_\text{os}(x,v,t)~.$
	Eq.~\eqref{s-rate} not only confirms the asserted bath-temperature $T_b= D/\gamma$ but also enables us to identify house-keeping entropy flux
	\begin{align}\label{ent-fl}
		\Phi_\text{hk} &:= \frac{q_\text{hk}}{T_b} = -\alpha \gamma\left( 1 - \frac{1}{T_b} \frac{\overline{D}}{\overline{\gamma}}\right) + \cdots~,
	\end{align} 
	and house-keeping entropy production
	\begin{align}\label{ent-pr}
		\Pi_\text{hk} :=\frac{1}{D_\alpha} \left\langle \left(\frac{\hat{J}_v^\text{ir}}{\hat{P}_\text{os}} \right)^2_t \right\rangle = \frac{ \alpha}{D}\left( \sqrt{\frac{\overline{D}}{\overline{\gamma}}}\gamma - \sqrt{\frac{\overline{\gamma}}{\overline{D}}}D \right)^2 + \cdots~.
	\end{align}
	The expression $\dot{S} =\Pi_\text{hk} - \Phi_\text{hk}$ implies that entropy flux and its production are not equal in OS, unlike in steady state, but their time-period averages are. The cyclic process viewpoint thus provides the physical picture where OS is sustained due to a dynamical interplay between entropy production in the system and heat exchange with the bath. 

	\begin{figure}[t]
		\includegraphics[width=\linewidth]{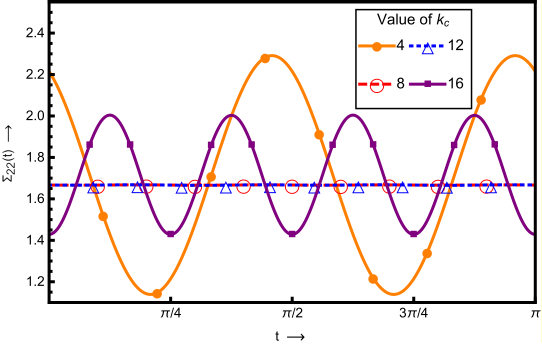}
		\caption{The asymptotic value of $\Sigma_{22} (t)$ have been plotted for different values of potential parameter  $k_0$. The viscous coefficient  $\gamma(t) = 3 + \sin(\omega t) + \cos(2 \omega t)$, noise strength $D(t) = 5 + 3\cos(\omega t)$, drive frequency $\omega = 4$ and $\alpha = 0.001$.}
		\label{resstat}
	\end{figure}

	In the leading order, we see that $P_{\text{os}}$ is time independent, its irreversible current is identically zero, the fluxes vanish and the entropy production rate is nil. The necessary and sufficient conditions for detailed balance~\cite{Kampen1957-1, Kampen1957-2, Graham1971, Gardiner1985} are thus satisfied, and the distribution is \`a la Boltzmann with an effective temperature $T_s^{(0)}= \overline{D}/\overline{\gamma}$. This is in contrast with $\alpha \rightarrow \infty$ case where the limiting distribution is far from equilibrium in spite of zero entropy production rate~\cite{Awasthi2022}. While OS tends to equilibrium when $\alpha \rightarrow 0$, the effective temperature is neither instantaneous bath-temperature nor its time average. For instance, the analytically solvable choice  \hbox{$D=D_0+D_1\cos(\omega t)$} and \hbox{$\gamma=\gamma_0+\gamma_1\cos(\omega t)$} leads to the relation
	\begin{equation}\label{tbbar}
		\overline{T}_b  - \frac{D_1}{\gamma_1}= \frac{\gamma_0}{\sqrt{\gamma_0^2 - \gamma_1^2}} \left( T_s^{(0)} - \frac{D_1}{\gamma_1} \right)~,
	\end{equation}
	which implies that the effective temperature can be greater or lesser than $\overline{T}_b$. If both $\overline{T}_b$ and $T_s^{(0)}$ are greater (lesser) than $D_1/\gamma_1$, then $T_s^{(0)}$ is lesser (greater) than $\overline{T}_b$.
		
	The results also suggest a mechanism to cool (or heat) a system with the help of a weakly interacting driven bath. Furthermore, they indicate, for instance, the possibility of maintaining a steady temperature indoors that keeps you cooler during day and warmer at night without consuming extra electricity~\cite{Raman2014}.

	\textsl{Case II}: 
	 When \hbox{$k_0 = n_0^2 \omega^2/4$} for some non-zero integer$~n_0$, we observe numerically that OS show significant time dependence at small $\alpha$ limit, and hence will be referred to as \textit{resonant states}. We plot as an example, the moment $\Sigma_{11}(t)$ in Fig.~\ref{resstat} for the choice $\gamma(t) = 3 + \sin(\omega t)+ \cos(2 \omega t)$, $D(t) = 5 + 3\cos(\omega t)$, $\omega = 4$ and $\alpha=0.001$. We observe that the resonant states occur only when $k_0=4\text{ or }16$, namely $k_0 = 4 n_0^2$, where $n_0=1\text{ or }2$ and are the only harmonics present in $\gamma$ or $D$. 
	
	The proposed perturbative method enables us to find analytically the moments of resonating states and explain the observed peculiarities. It is convenient to separate the constant term and $n_0$-th harmonic, where $n_0$ satisfies \hbox{$k_0 = n_0^2 \omega^2/4$}, and decompose the functions as 
	\begin{align}
		\gamma (t) =&  \gamma_0 +\gamma_+ e^{-i n_0 \omega t} + \gamma_- e^{i n_0 \omega t} + \gamma'(t)~, \\
		D (t) =& D_0 + D_+ e^{-i n_0 \omega t} +D_- e^{i n_0 \omega t} +D'(t)~,
	\end{align}
	where constants $\gamma_0$ and $D_0$ are real, $n_0$-th Fourier coefficients satisfy $\gamma_{\pm}=\gamma_{\mp}^*$ and $D_{\pm}=D_{\mp}^*$, and $\gamma'(t)$ and $D'(t)$ contain remaining modes. We find the leading-order moments to be
	\begin{align}
		\Sigma_{11}^{(0)} &=
		C_-^{(0)}  e^{ i n_0 \omega t} + C_0^{(0)} +C_+^{(0)}  e^{- i n_0 \omega t}~, \label{sig11-resm}\\
		\Sigma_{12}^{(0)} &=	\frac{i}{2}n_0 \omega \left( C_-^{(0)} e^{i n_0 \omega t} - C_+^{(0)} e^{- i n_0 \omega t} \right)~, \label{sig12-resm}\\
		\Sigma_{22}^{(0)} &=	-\frac{n_0^2 \omega^2}{4} \left( C_-^{(0)} e^{ i n_0 \omega t} - C_0^{(0)}  + C_+^{(0)} e^{- i n_0 \omega t}\right),\label{sig22-resm}
	\end{align}
	where 
	\begin{align} 
		C_\pm^{(0)} =& -\frac{2 D_\pm}{n_0^2 \omega^{2} \gamma_{0}}+\frac{C_0^{(0)} \gamma_{\pm}}{2 \gamma_{0}}~, \label{resc01m} \\ 
		C_0^{(0)} =& \frac{4 D_0 \gamma_0 - 2\left( D_+ \gamma_- + D_- \gamma_+\right)}{n_0^2 \omega^2 \left( \gamma_0^2 - \gamma_+\gamma_- \right)}~. \label{resc02m}
	\end{align}
	We see that the moments are time-dependent even in $\alpha \to 0$ limit and depend not only on time-averaged drives but also $n_0$-th Fourier components. If both $\gamma$ and $D$ do not contain $n_0$-th harmonic, OS belongs to {Case I}. The amplitudes of $\Sigma_{11}$ and $\Sigma_{12}$ decrease with increasing $n_0$, a trend noticed earlier in Fig.~\ref{resstat}. Interestingly, we further find that the time-period of OS at leading order is $T/n_0$. The result suggests a mechanism to activate a system in a higher-harmonic mode of the weakly interacting bath, and may have immense potential for pragmatic utility. Furthermore, we provide the moments to first-order and their numerical verification in SM~\cite{supp}. We note that the first-order corrections revert the periodicity of OS back to drive period$~T$.
		
	In spite of resonant states being time-dependent in the limit $\alpha \to 0$, we find as in {Case I} that the work done, the irreversible current, the house-keeping fluxes and entropy production in OS vanish. Hence the system in resonant OS is also energetically isolated with constant energy$~E$. We find $E = 2 k_0  C_0^{(0)}$, where $C_0^{(0)}$ is given by Eq.~\eqref{resc02m}, though both kinetic energy$~\langle E_k \rangle=\Sigma_{22}/2$ and potential energy $\langle U \rangle = k_0\Sigma_{11}/2$ are time dependent. Both change in synchrony by keeping the sum $\langle E_k\rangle +\langle U \rangle$ fixed, which can be verified explicitly using Eqs.~(\ref{sig11-resm}, \ref{sig22-resm}). While it may be hard to imagine an isolated system being in a time-dependent state, it does so by a perpetual exchange of energy between its position and velocity degrees of freedom. The maneuver of this exchange is encoded in the correlation function$~\Sigma_{12}$ given by Eq.~\eqref{sig12-resm}.

	\begin{figure}[t]
		\includegraphics[width=\linewidth]{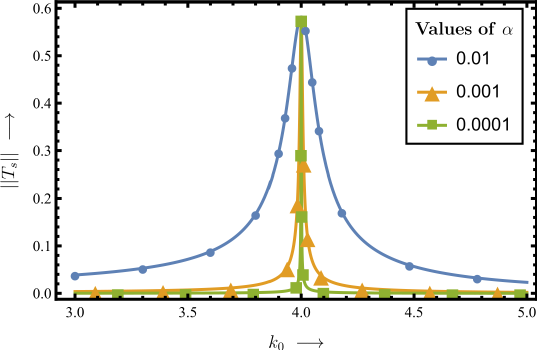}
		\caption{The amplitudes of oscillations of the system temperature $||T_s(t)||$ have been plotted for different values of potential parameter  $k_0$. The viscous coefficient  $\gamma(t) = 3 + \sin(\omega t) + \cos(2 \omega t)$, noise strength $D(t) = 5 + 3\cos(\omega t)$, drive frequency $\omega = 4$ and $\alpha = \lbrace0.01, 0.001, 0.0001\rbrace$.}
		\label{resstat4}
	\end{figure}

	Since resonant states emerge at fine-tuned condition, it may appear that they are impractical to realize, sustain and manipulate experimentally. The first-order corrections $\Sigma_{ij}^{(1)}$ in fact tell us that when $k_0$ deviates from $n_0^2 \omega^2/4$ by an amount $\Delta k \ll n_0 \omega^2$, then the contribution from $n_0$-th mode of the drives dominates over the rest. Thus operating the system at small non-zero $\alpha$, either by design or due to inevitable influence of the bath, is sufficient to maintain it in an OS showing features of resonant states. To illustrate the non-equilibrium behavior in the vicinity of resonant points for small$~\alpha$, we plot in Fig.~\ref{resstat4} the amplitude of oscillating system temperature$~T_s$ as we sweep$~k_0$, for the choice  $\gamma(t) = 3 + \sin(\omega t) + \cos(2 \omega t)$, $D(t) = 5 + 3\cos(\omega t)$ and $\omega = 4$. The amplitude dies down as $k_0$ deviates from $4$, thus signaling equilibrium. Moreover, the range of non-equilibrium behavior sharply falls off with decreasing$~\alpha$.

	The OS behavior in the limit $\alpha \to 0$ that we report is not restricted to harmonic potential alone. Both equilibrium and resonant limits continue to exhibit the same qualitative features when time-independent anharmonic terms are added. The resonant condition gets modified and can be systematically evaluated in anharmonic expansion.  When either harmonic or anharmonic terms become time dependent, then the leading behavior is no longer equilibrium even in the generic case. The proposed perturbative method though can be employed to analyze all these cases as detailed in the SM~\cite{supp}.

	%
	%
	
	\textsl{Conclusions.--}Low viscous physics of Langevin systems is relatively difficult to probe due to slow convergence and the singular nature of the limit. We have proposed a method based on singular perturbation and Floquet theories that is suitable to determine OS distributions of driven Langevin systems for low viscous drives. The proposed method is extensively verified with exact numerical calculations and simulations for determining OS of a driven under-damped Brownian particle moving in one-dimensional space in harmonic and quartic (driven) potentials. The method is in principle applicable to any driven (interacting multi-particle) Langevin systems non-perturbatively or to any order of perturbation. It can also be extended to Floquet quantum systems.  
	
	We find that the low viscous OS fall into two classes, distinguishable by a resonance condition, each with strikingly different physical features. In vanishing viscosity limit, OS belonging to the generic class tends to an equilibrium state with an effective temperature that can be tuned by viscous and thermal drives. This feature can be expected even in interacting systems and thus opening up possibilities for practical applications including cooling a system using a periodic time-dependent environment. For the fine-tuned class, when viscosity vanishes and the system is isolated from its environment, the OS continue to exhibit non-equilibrium properties where internal degrees resonate with one other in fractional time-period. This feature, expected to show up in interacting systems around fine-tuned parameters of the potentials, is indeed promising for various applications including higher harmonic activation~\cite{Crim2008, Ulasevich2020}. It would be interesting to come up with drive protocols for the environment to engineer resonant OS on systems (for example nano-robots~\cite{Rigatos2009}) required for some specific purpose.

	\bibliography{RefShakul4_PRL}

\end{document}


\title{Periodically driven thermodynamic systems under vanishingly small viscous drives\\ --- Supplemental Materials ---}
	\author{Shakul Awasthi}
	\email{shakul23@kias.re.kr}
	\affiliation{School of Physics, Korea Institute for Advanced Study, Seoul 02455, Korea}
	\author{Sreedhar B. Dutta}
	\email{sbdutta@iisertvm.ac.in}
	\affiliation{School of Physics, Indian Institute of Science Education and Research Thiruvananthapuram, Thiruvananthapuram 695551, India}
	
	\date{\today}

	\maketitle
	
	\newcommand{\bee}{\begin{equation}}
		\newcommand{\eee}{\end{equation}}
	\newcommand{\tm}{(t)}
	\newcommand{\gm}{\gamma}
	\newcommand{\dbar}{d\hspace*{-0.08em}\bar{}\hspace*{0.1em}}
	\newcommand{\haf}{\frac{1}{2}}
	\newcommand{\x}{\mathbf{x}}
	
	%
	%
	{	
		\hypersetup{linkcolor=black}
		\tableofcontents
	}
	
	\section{Perturbative scheme to determine OS of driven Brownian motion in time-dependent harmonic potentials}
	A Brownian particle that is subjected to periodic drives can reach a stable large-time non-equilibrium state called oscillating state (OS), whose distribution$~P_{os}^{(0)}(x,v,t)$ can be obtained from the asymptotic solution of its Fokker-Planck equation. When any parameter of the Brownian particle is made time-dependent with periodicity$~T$, we refer to it as a drive. In case of driven Brownian particle in harmonic potential drive, the distribution$~P_{os}^{(0)}(x,v,t)$ of OS is Gaussian with zero mean and time-periodic covariance matrix $\Sigma$~\cite{Awasthi2020}, whose matrix elements, $\Sigma_{11}=\left\langle x^{2}\right\rangle$, $\Sigma_{12}=\Sigma_{21}=\langle x v\rangle$ and $\Sigma_{22}=\left\langle v^{2}\right\rangle$, satisfy the equations 
	\begin{eqnarray}\label{mom-2-as}
		\frac{d}{dt}\Sigma_{11} &=& 2 \Sigma_{12}~, \nonumber \\
		\frac{d}{dt}\Sigma_{12} &=&-k \Sigma_{11} -\gamma_\alpha \Sigma_{12} + \Sigma_{22}~,\nonumber \\
		\frac{d}{dt}\Sigma_{22} &=& -2k \Sigma_{12} -2 \gamma_\alpha\Sigma_{22} +2D_\alpha~.
	\end{eqnarray}
	Here $k, \gamma_{\alpha}$ and $D_{\alpha}$ denote potential, viscous and noise drives, respectively. In order to monitor the viscous effects, it is convenient to consider a one-parameter extensions of viscosity $\gamma(t)\to\gamma_\alpha(t)=\alpha\gamma(t)$, and noise strength $D(t)\to D_\alpha(t)=\alpha D(t)$, where $\alpha$ is a non-negative real number. This extension keeps the bath temperature $T_b(t)~=~D(t)/\gamma(t)$ independent of $\alpha$ and isolates the effect of viscous drives. 

	The moments$~\Sigma_{ij}(t) = \sum_{n=0}^{\infty} \alpha^n \Sigma_{ij}^{(n)}(t)$ need to be determined perturbatively. Our strategy is to find arbitrary solution and impose periodicity to fix the arbitrary constants. But imposing periodicity on the leading terms$~\Sigma_{ij}^{(0)}(t)$ does not fix all the constants$~\Sigma_{ij}^{(0)}(0)$, which is a signature of singular limit. The first-order arbitrary solution$~\Sigma_{ij}^{(1)}(t)$ also contains arbitrary constants$~\Sigma_{ij}^{(1)}(0)$ in its complementary solution and the undetermined constants$~\Sigma_{ij}^{(0)}(0)$ in its particular solution. Now when we impose the periodicity condition at first-order, we find all$~\Sigma_{ij}^{(0)}(0)$ get fixed while some of$~\Sigma_{ij}^{(1)}(0)$ are undetermined as expected. In general periodicity at $(n+1)$-th order determines the moments of OS to $n$-th order.
	
	 The formal solution of Eq.~\eqref{mom-2-as}, written in terms of vector $\mathbf{X}_{2}=\left[\Sigma_{11}, \Sigma_{12}, \Sigma_{22}\right]^{T}$, is of the form
	\begin{equation}\label{mom-2-sol}
		\mathbf{X}_{2}(t)=K(t, 0) \mathbf{X}_{2}(0)+\int_{0}^{t} d s K(t, s) \mathbf{b}_\alpha(s)
	\end{equation}
	where$~K(t,s)$ is the evolution operator and column vector $\mathbf{b}_\alpha(t) = \left[ 0,0,2 D_\alpha\right]^{T}$. Here the superscript$~T$ denotes {\it transpose} operation.
	
 In order to obtain the OS moments~$\mathbf{X}_{2}(t)$, instead of taking the~$t \to \infty$ limit, we can also invoke Floquet theory and choose specific value of the initial conditions~$ \mathbf{X}_2(0)$ that renders the arbitrary solution~$ \mathbf{X}_2(t)$ $T$-periodic. This choice can be fixed by essentially demanding that $\mathbf{X}_{2}(T) = \mathbf{X}_{2}(0)$, namely
	\begin{equation}\label{x2-ini}
	({1 - K_2(T,0) })\mathbf{X}_{2}(0) = \int_{0}^{T} ds K_2(T,s)\mathbf{b}_\alpha(s)~.
	\end{equation}
	At $\alpha =0$, the matrix $[{1 - K_2(T,0) }]$ becomes singular and, furthermore, the integral on right-hand side of the Eq.~\eqref{x2-ini} vanishes. Hence $\mathbf{X}_{2}(0)$ cannot be completely determined in $\alpha \to 0$ limit, which is a consequence of the singular nature of this limit. It may appear that the perturbative expansion 
	\begin{align}
		\mathbf{X}_2(t) = \sum_{n=0}^{\infty} \alpha^n \mathbf{X}_2^{(n)}(t)
		= \left[ \Sigma_{11}^{(0)},\,\Sigma_{12}^{(0)},\,\Sigma_{22}^{(0)} \right]^T + \alpha \left[ \Sigma_{11}^{(1)},\,\Sigma_{12}^{(1)},\,\Sigma_{22}^{(1)} \right]^T + \cdots~,
	\end{align}
cannot be determined, since even the leading term of the initial condition
	\begin{equation}
		\mathbf{X}_{2}(0) = \mathbf{X}_{2}^{(0)}(0) + \alpha \mathbf{X}_{2}^{(1)}(0) + \alpha^2 \mathbf{X}_{2}^{(2)}(0) + \cdots~,
	\end{equation}
is not completely fixed. But, in fact, the first-order term of Eq.~\eqref{x2-ini} actually allows us to fix the zeroth-order$~\mathbf{X}_{2}^{(0)}(0)$ completely. In general, we can determine the coefficients $\mathbf{X}_{2}^{(n)}(0)$ to any order$~n$ by demanding periodicity of 
$\mathbf{X}_2(t)$ to order$~(n+1)$. 

The evolution operator can be written as $K(t,s)=M(t)M^{-1}(s)$, where$~M(t)$ is the fundamental matrix constructed from the solutions of Eqs.~\eqref{mom-2-as}. Hence we can obtain the perturbative expansion of $K(t,s)$ or $M(t)$ from that of the solutions, and vice versa. Note that the set of Eqs.~\eqref{mom-2-as} is essentially a third-order differential equation. Its solutions, and in turn the matrices can also be obtained, in terms of solutions of the second-order Hill equation 
	\begin{equation}\label{hill-eq-0}
		\frac{d^2u}{dt^2}  + \nu_\alpha u =0~,\quad \nu_\alpha = k  - \frac{1}{2}\dot{\gamma_\alpha} - \frac{1}{4}\gamma_\alpha^2~,
	\end{equation}
by following the procedure as detailed in Ref.~\cite{Awasthi2020}. Furthermore, the Floquet exponents$~\mu_{\alpha}$ associated with this equation also determine the existence condition of OS~\cite{Awasthi2020}, given by 
	\begin{equation}\label{stab-cond}
		|\operatorname{Re}(\mu_\alpha)|<\frac{1}{2} \overline{\gamma}_\alpha,
	\end{equation}
	where $\overline{\gamma}_\alpha$ denotes the average of $\gamma_\alpha(t)$ over a time-period$~T$. It is evident that the one-parameter extension of the stability condition \eqref {stab-cond} is violated in the limit $\alpha \rightarrow 0$, if the Floquet exponent $\operatorname{Re}\left(\mu_0\right) \neq 0$. Therefore, a necessary but not sufficient condition for the drive to maintain the stability of OS is $R e\left(\mu_0\right)=0$. Moreover, for smooth drives the Floquet exponents in general will have a series expansion $\mu_\alpha=\mu_0+\alpha \mu^{(1)}+\alpha^2 \mu^{(2)}+\cdots$, and hence the condition $\left|\operatorname{Re}\left(\mu^{(1)}\right)\right|<\bar{\gamma}/2$ should also hold to ensure OS stability.

	The perturbative solution 
		\begin{equation}\label{hill-eq-alpha}
		u_\alpha=\sum_{n=0}^{\infty} \alpha^n u^{(n)}~,
	\end{equation}
can be determined by solving the hierarchy of equations obtained at each order from Eq.~\eqref{hill-eq-0}. The leading-order equation is another Hill equation with $\nu_\alpha \to k$, given by
	\begin{equation}\label{hill-alp-0}
		\ddot{u}^{(0)}+k u^{(0)}=0~,
	\end{equation}
while the other components satisfy inhomogeneous equations of the form
	\begin{equation}\label{hill-alp-n}
		\ddot{u}^{(n)}+k u^{(n)}=f^{(n)},
	\end{equation}
	where
	\begin{equation}
		\begin{aligned}
			f^{(1)} &=\frac{1}{2} \dot{\gamma} u^{(0)}, \\
			f^{(n)} &=\frac{1}{2} \dot{\gamma} u^{(n-1)}+\frac{1}{4} \gamma^2 u^{(n-2)}, \quad \text { for } n \geq 2~.
		\end{aligned}
	\end{equation}
	Suppose we choose the two independent solutions of Eq.~\eqref{hill-eq-0} to be $u_{\alpha 1}(t)$ and $u_{\alpha 2}(t)$, satisfying the initial conditions $u_{\alpha 1}(0)=1, u_{\alpha 2}(0)=0, \dot{u}_{\alpha 1}(0)=0$ and $\dot{u}_{\alpha 2}(0)=1$, while those of Eq.~\eqref{hill-alp-n} to be $u_1^{(0)}(t)$ and $u_2^{(0)}(t)$, satisfying the initial conditions $u_1^{(0)}(0)=1$, $u_2^{(0)}(0)=0, \dot{u}_1^{(0)}(0)=0$ and $\dot{u}_2^{(0)}(0)=1$. Then the initial conditions for other components are fixed: $u_1^{(n)}(0)=0, u_2^{(n)}(0)=0, \dot{u}_1^{(n)}(0)=0$ and $\dot{u}_2^{(n)}(0)=$ 0 , for $n \geq 1$. Hence the two independent solutions of Eq.~\eqref{hill-eq-0} for any given $n$ are
	\begin{equation}\label{hig-ord-hill}
		u_k^{(n)}(t)=-\epsilon_{i j} u_i^{(0)}(t) \int_0^t d t^{\prime} u_j^{(0)}\left(t^{\prime}\right) f_k^{(n)}\left(t^{\prime}\right),
	\end{equation}
	where the summation over repeated indices is implied, the indices $i, j$ and $k$ run over 1 and 2, the matrix $\epsilon$ is the Levi-Civita matrix with $\epsilon_{12}=1$, and the two functions $f_k^{(n)}$ are defined from $f^{(n)}$ associated to each solution.

	We can now, in principle, obtain the series expansion of the Floquet exponent $\mu_\alpha$ from the linear relations
	\begin{equation}\label{u-alpha-i}
		u_{\alpha i}(T)=\Phi_{i j}(T) u_{\alpha j}(0)~,
	\end{equation}
	which enable us to construct the matrix $\Phi$ whose eigenvalues are $\exp \left(\pm \mu_\alpha T\right)$. We will instead evaluate this matrix only to sub-leading order since that is sufficient to corroborate the stability of OS. It follows from Eqs.~(\ref{hig-ord-hill},~\ref{u-alpha-i}) that
	\begin{equation}
		u_i^{(1)}(T)=T B_{i j}(T) \epsilon_{j k} u_k^{(0)}(T)~,
	\end{equation}
	where 
	\begin{equation}
		B_{i j}(T):=\frac{1}{2 T} \int_0^T d t^{\prime} \dot{\gamma}\left(t^{\prime}\right) u_i^{(0)}\left(t^{\prime}\right) u_j^{(0)}\left(t^{\prime}\right) ~,
	\end{equation}
which can be considered as elements of a symmetric matrix$~B$.
	Hence to first order in $\alpha$, the matrix
	\begin{equation}\label{phi-exp}
		\Phi=\Phi^{(0)}+\alpha T B \epsilon \Phi^{(0)}+\cdots~,
	\end{equation}
	where $\Phi^{(0)}$ is defined from the relations
	\begin{equation}
		u_i^{(0)}(T)=\Phi_{i j}^{(0)}(T) u_j^{(0)}(0)~,
	\end{equation}
	and whose eigenvalues are $\exp \left(\pm \mu_0 T\right)$. It is straightforward to show that Eq.~\eqref{phi-exp} leads to
	\begin{equation}\label{mu-alp-1}
		\mu_\alpha=\mu_0+\alpha \frac{\operatorname{Tr}\left(B \epsilon \Phi^{(0)}\right)}{\left(e^{\mu_0 T}-e^{-\mu_0 T}\right)}+\cdots~,
	\end{equation}
	Since $\mu_0 = 0$ is a necessary condition, it may appear that the sub-leading term $\mu_1$ blows up. This however is not the case, as we will see explicitly in Sec.~\ref{small-alph-k0-sec} for constant$~k$.
	
	The fundamental matrix $M=\sum \alpha^n M^{(n)}$, which can be written in terms of the solution $u_{\alpha}$~\cite{Awasthi2020}, can now be Taylor expanded straightforwardly.  The expression for zeroth-order is 
	\begin{align}\label{M0-mat}
		M^{(0)} =& 
		\begin{bmatrix}
			(u_1^{(0)})^2 & u_1^{(0)} u_2^{(0)} & (u_2^{(0)})^2 \\
			\dot{u_1^{(0)}} u_1^{(0)} & \frac{1}{2} \left(\dot{u_2^{(0)}} u_1^{(0)}+\dot{u_1^{(0)}} u_2^{(0)}\right) &
			\dot{u_2^{(0)}} u_2^{(0)} \\
			(\dot{u_1^{(0)}})^2 & \dot{u_2^{(0)}} \dot{u_1^{(0)}} & (\dot{u_2^{(0)}})^2 
		\end{bmatrix}~,
	\end{align}
	where $u_1^{(0)}$ and $u_2^{(0)}$ are the two independent solutions of the Hill equation \eqref{hill-alp-0} and depend only on the harmonic potential strength~$k(t)$. While, the expression for first-order term is
	\begin{align} \label{M1-mat}
		M^{(1)} = &\begin{bmatrix}
			\widetilde{M}_{11} & \widetilde{M}_{12} & \widetilde{M}_{13} \\
			\widetilde{M}_{21} &
			\widetilde{M}_{22} &\widetilde{M}_{23}\\
			\widetilde{M}_{31} &\widetilde{M}_{32}  & \widetilde{M}_{33}
		\end{bmatrix} 
		- \Gamma M^{(0)}~,
	\end{align}
	where the quantities denoted by $\widetilde{M}$ are given by
	\begin{align}
		\widetilde{M}_{11} =& 2 u_1^{(0)} u_1^{(1)}~,\nonumber\\
		\widetilde{M}_{12} =& u_1^{(0)} u_2^{(1)}+u_1^{(1)} u_2^{(0)}~,\nonumber\\
		\widetilde{M}_{13} =& 2 u_2^{(0)} u_2^{(1)}~,\nonumber\\
		\widetilde{M}_{21} =& \dot{u_1^{(0)}} u_1^{(1)}-\frac{1}{2} \gamma  u_1^{(0)} u_1^{(0)}+\dot{u_1^{(1)}} u_1^{(0)}~,\nonumber\\
		\widetilde{M}_{22} =& \frac{\dot{u_2^{(0)}} u_1^{(1)}}{2}-\frac{\gamma  u_1^{(0)} u_2^{(0)}}{4}+\frac{\dot{u_1^{(0)}}
			u_2^{(1)}}{2}+\frac{\dot{u_1^{(1)}} u_2^{(0)}}{2}-\frac{\gamma  u_2^{(0)} u_1^{(0)}}{4}+\frac{\dot{u_2^{(1)}}
			u_1^{(0)}}{2}~,\nonumber\\
		\widetilde{M}_{23} =& \dot{u_2^{(0)}} u_2^{(1)}-\frac{\gamma  u_2^{(0)} u_2^{(0)}}{2}+\dot{u_2^{(1)}} u_2^{(0)}~,\nonumber\\
		\widetilde{M}_{31} =& 2 \dot{u_1^{(0)}} \dot{u_1^{(1)}}-\gamma  u_1^{(0)} \dot{u_1^{(0)}}~,\nonumber\\
		\widetilde{M}_{32} =& -\frac{1}{2} \gamma  \dot{u_2^{(0)}}
		u_1^{(0)}+\dot{u_2^{(0)}} \dot{u_1^{(1)}}-\frac{1}{2} \gamma  \dot{u_1^{(0)}} u_2^{(0)}+\dot{u_1^{(0)}}
		\dot{u_2^{(1)}}~,\nonumber \\
		\widetilde{M}_{33} =& 2 \dot{u_2^{(0)}} \dot{u_2^{(1)}}-\gamma  \dot{u_2^{(0)}} u_2^{(0)}~,\label{mtilde}
	\end{align} 
	and $u_1^{(1)}$ and $u_2^{(1)}$ can be found using the Eq.~\eqref{hill-alp-n}. Henceforth, for simplicity of notation, we use $u_\pm$ to denote the solutions~$u_1$ and $u_2$. 
	
	We can similarly construct higher coefficients of the fundamental matrix and find higher-order corrections to the moments, as will be done in the remainder.

	\section{Moments of driven Brownian particle in time-independent harmonic potential} \label{small-alph-k0-sec}
	
	The perturbative solution for small viscous drives is analytically obtainable when the zeroth-order Hill equation~\eqref{hill-alp-0} is exactly solvable. This is, in general, not the case for any time-periodic function~$k(t)$. In this section, we restrict $k=k_0$, where $k_0$ is constant, and explicitly write down second moments of OS to first-order in $\alpha$. We will later, in Sec.~\ref{sml-apl-kt}, consider a less restrictive case with $k=k_0 +\delta k(t)$, where $\delta k(t)$ is weak time-dependent perturbation. 
	
	The two solutions of the Hill equation \eqref{hill-alp-0} for constant$~k_0$ are 
	\begin{equation}\label{zero-ode-sol}
		u^{(0)}_{\pm} = e^{\pm i \sqrt{k_0} t }~.
	\end{equation}
	Thus the leading-order Floquet exponent$~\mu_0$ is purely imaginary for$~k_0~>~0$, and one of the necessary conditions $R e\left(\mu_0\right)=0$ for existence of OS is fulfilled. It is convenient to Fourier expand the drives,
		\begin{equation}\label{frsr-1}
		\gamma(t) = \sum_{n=-\infty}^{\infty} \gamma_n e^{- i n \omega t }~,\quad D(t) = \sum_{n=-\infty}^{\infty} D_n e^{- i n \omega t }~,
	\end{equation} 
	where $\gamma_n^*=\gamma_{-n}$ and $D_n^*=D_{-n}$. The higher-order corrections can be evaluated using Eq.~\eqref{hig-ord-hill}. We explicitly obtain the solution of Eq.~\eqref{hill-eq-0} to $O(\alpha^2)$, which is given by the expression
	\begin{equation} \label{sakcsol}
		u_{\pm} = e^{i \left( \pm \sqrt{k_0} \mp \alpha^2 \mu_2\right) t}\left(1 + \alpha P_{\pm}^{(1)}(t) + \alpha^2  P_{\pm}^{(2)}(t)\right)~,
	\end{equation}
	where the functions
	\begin{align}\label{hill-p1}
		P_{\pm}^{(1)} =& \sum_{n = -\infty}^{\infty}  \frac{\pm i \overline{\delta}_{n,0} \gamma_n}{2 n \omega \pm 2 \sqrt{k_0}}e^{i n \omega t}~, \\ \label{hill-p2}
		P_{\pm}^{(2)} =& \sum_{n = -\infty}^{\infty} \overline{\delta}_{n,0}\left(\pm \frac{ \gamma_n \gamma_0}{2 n \omega (2 \sqrt{k_0} - n \omega)} \pm  \sum_{m=-\infty}^{\infty}  \frac{ \overline{\delta}_{n,m} \overline{\delta}_{m,0} \gamma_m \gamma_{n-m}}{4 n \omega (2 \sqrt{k_0} \mp n \omega(n-m))} \right)~,
	\end{align}
	are purely periodic,  the quantity 
	\begin{equation}
		\mu_2 = \sum_{n=-\infty}^{\infty} \frac{|\gamma_n|^2}{4(2 \sqrt{k_0} + n \omega)}~,
	\end{equation}
is a real constant, and $\overline{\delta}_{n,m} = 1 - \delta_{n,m}$ is related to Kronecker delta $\delta_{n,m}$. The solutions \eqref{sakcsol} are written in the pseudo-periodic form, clearly indicating that Floquet exponents are purely imaginary to $O(\alpha^2)$ for $k_0>0$ and thus respecting the stability condition of OS. In this section, the solutions \eqref{sakcsol} are obtained assuming $k_0 \ne n_0^2\omega^2/4$, for any integer$~n_0$. The special cases, when $k_0 = n_0^2\omega^2/4$, for some integer$~n_0$, will be discussed later in Sec.~\ref{Para-res}. 

 It is straightforward to evaluate the matrix $M$ using Eqs.~(\ref{M0-mat},~\ref{M1-mat}), Fourier expand the vector $\mathbf{b}_{\alpha}(s)$ using \eqref{frsr-1}, and substitute both in Eq.~\eqref{mom-2-sol} to obtain the moments to $O(\alpha)$. The unknown constant vector $C := M^{-1}(0)\mathbf{X}_{2}(0)$, or equivalently its components $C_n = \sum_m \alpha^m C_n^{(m)}$, can be obtained to first order by demanding periodicity of  $\mathbf{X}_{2}(t)$ to $O(\alpha^2)$. We find the leading term to be
	\begin{equation}\label{2-mom-0}
		\mathbf{X}_{2}^{(0)} =
		\begin{bmatrix}
			C_1^{(0)}  e^{2 i \sqrt{k_0} t} + C_2^{(0)} +C_3^{(0)}  e^{-2 i \sqrt{k_0} t} \\
			i \sqrt{k_0} \left( C_1^{(0)} e^{2 i \sqrt{k_0}t} - C_3^{(0)} e^{-2 i \sqrt{k_0} t} \right)\\
			k_0 \left( - C_1^{(0)} e^{2 i \sqrt{k c} t} + C_2^{(0)}   - C_3^{(0)} e^{-2 i \sqrt{k c} t}\right) \\
		\end{bmatrix}~,
	\end{equation}
	where $C_3^{(0)}= C_1^{(0)*}$ and $C_2^{(0)}= C_2^{(0)*}$. Periodicity at leading-order fixes only two constants, $C_1^{(0)} = C_3^{(0)} = 0$, leaving behind $C_2^{(0)}$ undetermined. We then use Eqs.~\eqref{mtilde}, which require the first-order terms$~u_{\pm}^{(1)}$, to obtain moments to $O(\alpha)$. In fact, we only need $\Sigma_{11}^{(1)}$, to fix $C_{2}^{(0)}$, which is found to be
	\begin{align}
		\Sigma_{11}^{(1)} =& 	C_1^{(1)}  e^{2 i \sqrt{k_0} t} + C_2^{(1)} +C_3^{(1)}  e^{-2 i \sqrt{k_0} t} + \left(\frac{D_0}{k_0} - \gamma_0 C_2^{(0)}\right) t + \sum_{n\neq 0} \bigg[ \frac{i}{n \omega k_0}\left( -D_n + C_2^{(0)} k_0 \gamma_n  \right) +\nonumber \\ & \left( \frac{-2 i \sqrt{k_0} \gamma_n}{n \omega (n \omega + 2 \sqrt{k_0})} \right) e^{-i(n \omega + 2 \sqrt{k_0})t} + \left( \frac{2 i \sqrt{k_0} \gamma_n}{n \omega (n \omega + 2 \sqrt{k_0})} \right) e^{-i(n \omega - 2 \sqrt{k_0})t} + \left(\frac{-4 i (D_n - C_2^{(0)} k_0 \gamma_n)}{n \omega (n^2 \omega^2 - 4 k_0)}\right) e^{-i n \omega t} +  \nonumber \\ &\frac{i}{4} \left( \frac{D_0}{k_0^{3/2} }+ \frac{2 D_n}{n \omega + 2 k_0^{3/2}}  \right) e^{2 i \sqrt{k_0} t} + \frac{-i}{4} \left( \frac{D_0}{k_0^{3/2} }- \frac{2 D_n}{n \omega + 2 k_0^{3/2}}  \right) e^{-2 i \sqrt{k_0} t} \bigg]
	\end{align}
	Periodicity of $\Sigma_{11}^{(1)}$ implies
	\begin{equation}
		C_1^{(1)} = C_3^{(1)} = 0, \quad C_2^{(0)} = \frac{D_0}{k_0 \gamma_0} =  \frac{1}{k_0} \frac{\overline{D}}{ \overline{\gamma}}~,
	\end{equation}
	where the overline denotes time average over a period. Similarly, we demand periodicity of $\Sigma_{11}^{(2)}$ and fix $C_2^{(1)}$. We thus obtain the first-order corrections to be 
	\begin{align}\label{sig111}
		\Sigma_{11}^{(1)} =&  C_2^{(1)} + \frac{4 i}{\omega} \sum_{n=-\infty}^{\infty} 
		\frac{r_n}{n}  \frac{\overline{\delta}_{n,0} e^{- i n \omega t}}{n^2 \omega^2 - 4 k_0}~,\\ \label{sig121}
		\Sigma_{12}^{(1)} =& \sum_{n=-\infty}^{\infty} \frac{2 \overline{\delta}_{n,0}  r_n }{n^2 \omega^2 - 4 k_0 } e^{-i n \omega t}~,\\ \label{sig221}
		\Sigma_{22}^{(1)} =&  k_0 C_2^{(1)} + \sum_{n=-\infty}^{\infty}\left( \frac{\overline{\delta}_{n,0} 2 i r_n}{n \omega}\right) \left( \frac{ n^2 \omega^2 - 2 k_0}{ n^2 \omega^2 - 4 k_0} \right) e^{-i n \omega t}~,		
	\end{align}
	where
	\begin{align} \label{rnval}
		r_n = &  \frac{D_0}{\gamma_0}\gamma_n  -  D_n~, \\ \label{c0val}
		C_2^{(1)} = &  \sum_{n=-\infty}^{\infty} \overline{\delta}_{n,0} \frac{-2 i (n^2 \omega^2 - 2 k_0) D_n \gamma_{-n}}{n \omega (n^2 \omega^2 - 4 k_0) k_0 \gamma_0} ~.
	\end{align}
	
	\noindent \textbf{An illustrative example}\\
	
	\begin{figure}[t]
		\includegraphics[width=0.56\linewidth]{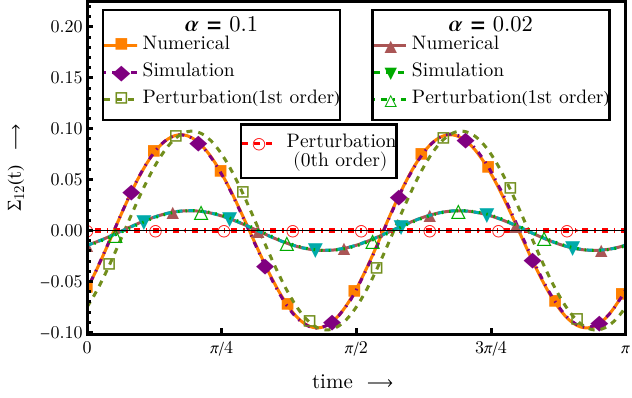}
		\caption{The asymptotic value of $\Sigma_{12} (t)$ has been plotted for two time periods. The viscous coefficient  $\gamma(t) = 2 + \sin(\omega t) $,  noise strength $D(t) = 5 + 3\cos(\omega t) $, drive frequency $\omega = 4$ , potential parameter $k_0 = 2$ for $\alpha = \lbrace 0.1,0.02 \rbrace$.}
		\label{smlalp}
	\end{figure}	
	
	We plot as an example the moment $\Sigma_{12}(t)$ in Fig.~\ref{smlalp}, for the choice $\gamma(t) = 2 + \sin(\omega t) $, $D(t) = 5 + 3\cos(\omega t) $, $\omega = 4$, and  $k_0 = 2$. The first-order correction for this choice is $\Sigma_{12}^{(1)} = -(3/4) \cos(4 t) + (5/8) \sin(4 t)$, which can be calculated using Eq.~\eqref{sig121}. The plot compares exact numerical results, obtained by solving moments equation Eq.~\eqref{mom-2-as} for large times, with those obtained perturbatively to first order. The perturbative results increasingly agree with the exact numerical ones as $\alpha$ approaches zero. The results of the Langevin simulation, shown for reference, also coincide as expected with the asymptotic numerical results. The same trend is observed for the other two moments $\Sigma_{20}(t)$ and $\Sigma_{02}(t)$ too.

	\section{Parametric Resonance in driven Brownian motion}\label{Para-res}
	In this section, we will explicitly obtain the moments of OS, to $O(\alpha)$, for the special case $k_0= n_0^2\omega^2/4$, where $n_0$ is a non-zero integer. Note that the expressions in Eqs.~(\ref{sig111}-\ref{sig221}) and Eq.~\eqref{hill-p1} diverge and are not applicable to this case.

		\begin{figure}[t]
		\includegraphics[width=0.6\linewidth]{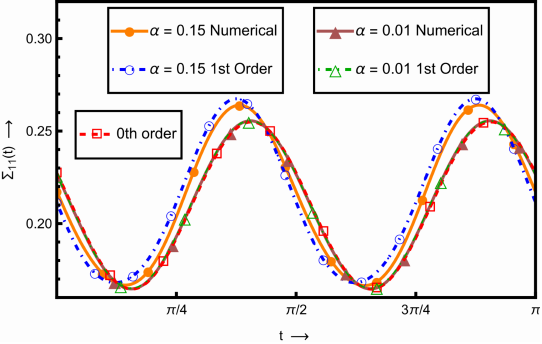}
		\caption{The asymptotic value of $\Sigma_{11} (t)$ has been plotted for two time periods. The viscous coefficient  $\gamma(t) = 6 + \cos(\omega t)+ 3 \sin(\omega t)$,  noise strength $D(t) = 5 + 2\sin(\omega t) + \sin(\omega t)$, drive frequency $\omega = 4$ , potential parameter $k_0 = 4$ for $\alpha = \lbrace 0.15,~0.01 \rbrace$.}
		\label{smlalpres}
	\end{figure}

	Let us assume that $k_0$ satisfies the special case for some integer$~n_0$. It is convenient to separate this mode in the drives, assuming it exists, and express them as 
	\begin{align}
		\gamma (t) =& \sum_{n=-\infty}^{\infty} \overline{\delta}_{n,n_0} \gamma_n e^{-i n \omega t} + \gamma_+ e^{-i n_0 \omega t} + \gamma_- e^{i n_0 \omega t} ~,\\
		D (t) =& \sum_{n=-\infty}^{\infty} \overline{\delta}_{n,n_0} D_n e^{-i n D t} + D_+ e^{-i n_0 \omega t} +D_- e^{i n_0 \omega t}~,
	\end{align} 
where $\gamma_{\pm}=\gamma_{\mp}^*$, $D_{\pm}=D_{\mp}^*$, $\gamma_{n}=\gamma_{-n}^*$ and $D_{n}=D_{-n}^*$, for all integers$~n \ne n_0$. Then the solutions of Eqs.~(\ref{hill-alp-0},~\ref{hill-alp-n}), to $O\left( \alpha \right)$, are
	\begin{align}\label{u-res}
		u_\pm = e^{\pm \frac{i}{2} n_0 \omega t} \left[ 1 + \alpha \left(\frac{\gamma_\pm t}{2} e^{\mp i  n_0 \omega t} \mp \frac{i \gamma_\mp}{4 n_0 \omega}e^{\pm i  n_0 \omega t} \mp \sum_{n=-\infty; n \neq n_0}^{\infty} \frac{i \gamma_n e^{-i n \omega t}}{2 (n_0 \mp n)\omega} \right) + O(\alpha^2) \right]~.
	\end{align} 
	These solutions are not in the pseudo-periodic form, but straightforward calculations using \eqref{mu-alp-1} lead to the Floquet exponents, whose real part vanishes for $k_0>0$, and confirm the stability of OS. Substituting the solutions in Eq.~\eqref{M0-mat} determines $M^{(0)}$ and gives, upon using the relation $\mathbf{X}_{2}^{(0)}=M^{(0)}C$, the leading-order moments
	\begin{align}
		\Sigma_{11}^{(0)} &=
		C_1^{(0)}  e^{ i n_0 \omega t} + C_2^{(0)} +C_3^{(0)}  e^{- i n_0 \omega t}~,\label{sig11-res} \\
		\Sigma_{12}^{(0)} &=	\frac{i}{2}n_0 \omega \left( C_1^{(0)} e^{i n_0 \omega t} - C_3^{(0)} e^{- i n_0\omega t} \right) \label{sig12-res}~,\\
		\Sigma_{22}^{(0)} &=	-\frac{n_0^2 \omega^2}{4} \left( C_1^{(0)} e^{ i n_0 \omega  t} - C_2^{(0)}  + C_3^{(0)} e^{- i n_0 \omega t}\right)~,
	\end{align}
where $C_3^{(0)}= C_1^{(0)*}$ and $C_2^{(0)}= C_2^{(0)*}$. These equations are of course similar to the earlier generic case. However, unlike there where periodicity at leading-order determined two of the three constants, in this special case none of the constants get determined at $O(\alpha)$ and are fixed only when periodicity is implemented at $O(\alpha)$. Thus we obtain  
	\begin{align} 
		C_1^{(0)} =& C_3^{(0)*}= -\frac{2 D_-}{n_0^{2} \omega^{2} \gamma_{0}}+\frac{C_2^{(0)} \gamma_{-}}{2 \gamma_{0}}~, \label{resc01} \\ 
		C_2^{(0)} =& \frac{4 D_0 \gamma_0 - 2\left( D_+ \gamma_- + D_- \gamma_+\right)}{n_0^2 \omega^2 \left( \gamma_0^2 - \gamma_+\gamma_- \right)} ~. \label{resc02}
	\end{align}
Similarly, we determine the $O(\alpha)$ corrections to the moments of OS in this special case too, having imposed periodicity to its arbitrary solution at $O(\alpha^2)$. We find
	\begin{align}\label{reso1}
		\mathbf{X}_{2}^{(1)} =   \sum_{n=\infty}^{\infty} \overline{\delta}_{n,n_0} \overline{\delta}_{n,0} \mathbf{P}_n e^{-i n \omega t} + \sum_{p=-2}^{2} \mathbf{Q}_p e^{-i p m \omega t}~,   
	\end{align}
	where the three elements of the column matrix $\mathbf{P}_n$ are 
	\begin{align}
		(\mathbf{P}_n)_{11} =& 
		\frac{i e_n}{n} + \frac{i C_3^{(0)} n_0 \gamma_{n-n_0}}{(n_0-n) n \omega} +  \frac{i C_1^{(0)} n_0 \gamma_{n+n_0}}{(n_0+n) n \omega}~,\\
		(\mathbf{P}_n)_{21} =& \frac{-2}{i n_0 \omega}\left( \frac{i e_n}{n_0} - \frac{i C_3^{(0)}  \gamma_{n-n_0}}{(n_0-n) \omega} +  \frac{i C_1^{(0)} \gamma_{n+n_0}}{(n_0+n)  \omega} \right)~,\\
		(\mathbf{P}_n)_{31} =& \frac{-4}{n_0^2 \omega^2}\left( \frac{- i (2n^2 -n_0^2) e_n}{n_0^2 n} - \frac{i C_3^{(0)} (2n-n_0) \gamma_{n-n_0}}{n(n-n_0) \omega} -  \frac{i C_1^{(0)}(2n+n_0) \gamma_{n+n_0}}{n(n_0+n) \omega} \right)~,
	\end{align}
	and 
	\begin{equation}
		e_n = \frac{-4 D_n + C_2^{(0)}n_0^2 \omega^2 \gamma_n}{(n^2 -n_0^2) \omega^3}~.
	\end{equation}
	The first-order correction, more specifically the first term on the right-hand side of Eq.~\eqref{reso1}, breaks the $T/n_0$-periodicity of the leading term and restores the original $T$-periodicity of the drives. We do not explicitly write down the terms$~Q_n$, which are cumbersome to express. However, we can verify these expressions using a numerical example.\\

		\noindent \textbf{An illustrative example}\\
		
	We have obtained all the moments analytically to first order and verified with exact numerical results for several choices.  For instance, the choice $\gamma(t) = 6 + \cos(\omega t)+ 3\sin(\omega t) $, $D(t) = 5 + 2\sin(\omega t) + \sin(2 \omega t) $, $\omega = 4$, and $k_0 = 4$ gives the expression for $\Sigma_{11}^{(1)}(t)$ as 
	\begin{align} 
		\Sigma_{11}^{(1)} &= \frac{589}{18304} - \frac{67}{572} \cos(4 t) - \frac{1537}{27456} \sin(4 t)  - \frac{577}{27456} \cos(8 t) - \frac{5}{2288} \sin(8 t) + \frac{5}{4576} \cos(12 t) - \frac{1}{384} \sin(12 t)~.
	\end{align}
	We plot the $\Sigma_{11}$ shown in Fig.~\ref{smlalpres} and confirm that the proposed perturbative method determines OS distributions with increasing accuracy even for resonant states as $\alpha$ decreases.

	\section{When perturbed by harmonic potential drive}\label{sml-apl-kt}
\begin{figure}[t!]
		\includegraphics[width=0.6\linewidth]{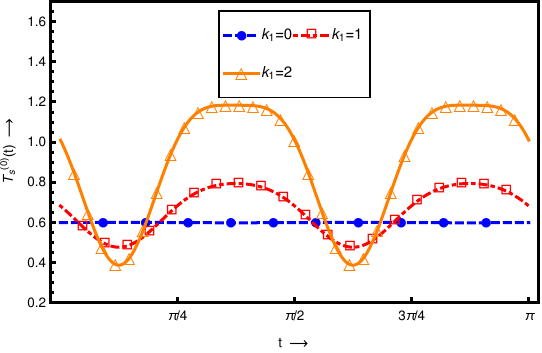}
		\caption{The kinetic temperature $T_s^{(0)} (t)$ has been plotted for two time periods. The viscous coefficient  $\gamma(t) = 5 + \cos(\omega t)$,  noise strength $D(t) = 3 + 2\sin(\omega t) + \sin(\omega t)$, drive frequency $\omega = 4$ , harmonic potential parameter $k(t) = 4 + \cos(\omega t)$  for $\alpha = 0.001$.}
		\label{Tskt}
	\end{figure}

	We have observed that OS distribution of driven Brownian motion in time-independent harmonic potential becomes equilibrium for the generic case in the limit $\alpha \to 0$. We can expect the leading behavior to be non-equilibrium for driven potentials. This can also be confirmed numerically by determining, say, kinetic temperature as shown in Fig.~\ref{Tskt}. As we increase the amplitude$~k_1$ of time-dependent part of the driven harmonic potential, the temperature departs from being a constant and attains an increasingly oscillatory behavior. Moreover, the time-averages change with increasing $k_1$ and take the values $\lbrace 0.60,\,0.66,\,0.90\rbrace$ for ${k_1~=~\lbrace0,\,1,\,2\rbrace}$ respectively. 
	
The perturbative scheme that we have proposed is equally applicable for time-dependent potentials. Hence, we can analyze the leading behavior of OS subjected to potential drives, which we will do now by choosing, for simplicity, harmonic drive of the form
	\begin{equation}\label{kt-form}
		k(t) = k_0 + k_1 \cos(\omega t + \phi_k)~,
	\end{equation}
where $\phi_k$ denotes the constant phase factor. We will further treat$~k_1$ perturbatively, and write the solutions of Hill equation~\eqref{hill-alp-0} in the form
	\begin{equation}\label{u01-sol-expa}
	u^{(0)}_{\pm} = u^{(0,0)}_{\pm} +  k_1 u^{(0,1)}_{\pm} + \cdots
\end{equation}	
The leading terms $u^{(0,0)}_{\pm}$ are the two independent solutions given by Eq.~\eqref{zero-ode-sol}, while the $O(k_1)$ correction 
	\begin{equation}\label{hill-sol-kt}
		u^{(0,1)}_\pm = \frac{k_1 e^{\pm \sqrt{k_0 t}}}{\omega^2 - 4 k_0} \left( \cos(\omega t + \phi_k) \mp \frac{2 i \sqrt{k_0}}{\omega} \sin(\omega t + \phi_k) \right)~.
	\end{equation}
	We play with the choice
	\begin{equation}
		D(t) = D_0 + D_1 \cos(\omega t),\quad \gamma(t) = \gamma_0 + \gamma_1 \cos(\omega t + \phi)~,
	\end{equation}
keeping a phase difference$~\phi$ between the two drives. Following the proposed perturbative scheme has resulted in the expressions,
	\begin{equation}\label{moments-kt}
\Sigma_{ij}^{(0)} = \Sigma_{ij}^{(0,0)} + k_1 \Sigma_{ij}^{(0,1)}  ~,
\end{equation}
where $\Sigma_{ij}^{(0,0)}$ are the same leading-order moments that we have obtained for $k=k_0$ case; and the $O(k_1)$ corrections to them are
	\begin{align}
		\Sigma_{11}^{(0,1)} &= \frac{1}{k_0} \frac{\overline{D}}{\overline{\gamma}}\left(\frac{\gamma_1 \cos(\phi-\phi_k)}{\gamma_0 \left( \omega^2 - 4 k_0 \right)} + \frac{2 \cos(\omega t + \phi_k)}{\omega^2 - 4 k_0}\right)~,\label{sig110-kt}\\ 
		\Sigma_{12}^{(0,1)} &= -\frac{1}{k_0} \frac{\overline{D}}{\overline{\gamma}}\frac{\omega \sin(\omega t + \phi_k)}{\omega^2 - 4 k_0}~, \label{sig120-kt}\\
		\Sigma_{22}^{(0,1)} &= \frac{\overline{D}}{\overline{\gamma}} \left(\frac{\gamma_1 \cos(\phi-\phi_k)}{\gamma_0 \left( \omega^2 - 4 k_0 \right)} -  \frac{2  \cos(\omega t + \phi_k)}{\omega^2-4 k_0} \right)~.\label{sig220-kt} 
	\end{align}
	These corrections due to potential drive will modify the thermodynamic characteristics of the OS. Note that the time-dependent part of noise strength $D(t)$ has no effect on the corrections.
	
	We can now explicitly find the $O(k_1)$ corrections to various thermodynamic quantities and analyze, among other properties, their time-averages and phase relationships with the input drives. We see the kinetic temperature $T_s = \Sigma_{22}$ is modified from $\overline{D}/\overline{\gamma}$ by the correction given in Eq.~\eqref{sig220-kt}. Thus $T_s$ at $O(\alpha^0)$ becomes time-dependent and its amplitude changes in synchrony with that of potential drive. Its average depends on the phase difference $(\phi-\phi_k)$ and shows maximum modification when this difference vanishes. The configuration temperature $T_c = k(t) \Sigma_{11}$ also becomes time-dependent and is given by
	\begin{align}
		T_c = \frac{D_0}{\gamma_0} & \left[ 1+\frac{ k_1 \gamma_1 \cos (\phi-\phi_k)}{\gamma_0 \left(\omega ^2- 4 k_0\right)} + 
		\frac{k_1\left(\omega ^2 - 2 k_0\right)}{k_0\left(\omega ^2 - 4 k_0\right)}\cos ( \omega t+\phi_k )\right]~,\label{Tc-kt} 
	\end{align}
showing similar amplitude and phase response to the drive as kinetic temperature. The difference between these temperatures, at $O(\alpha^0)$, turns out to be
	\begin{equation}\label{eq-part-kt}
		T_s - T_c = -\frac{k_1}{k_0} \frac{D_0}{\gamma_0}\frac{\omega ^2 \cos (\omega t+\phi_k)}{\left(\omega ^2 - 4 k_0\right)}~.
	\end{equation} 
This relation implies that the equipartition holds on an average and is violated otherwise. 

The correlations between position and velocity degrees of freedom are  non-zero at leading-order in the presence of potential drives, as can be seen from Eq.~\eqref{sig120-kt}. The correlation function and equipartition violation are directly proportional to each other but are out of phase by $\pi/2$, implying that the violation is at its minimum when the correlation is at its maximum, and vice-versa. 

Potential drives do work on the system, and here we find the rate of work done 
\begin{eqnarray}
	\left\langle \frac{\dbar W}{dt} \right \rangle = -\frac{k_1}{2}\frac{\overline{D}}{\overline{\gamma}} \frac{\omega}{k_0} \sin(\omega t + \phi_k)~,
\end{eqnarray}
is non-zero, though its average vanishes. This rate is in phase with the correlations, and increases proportionally with the drive frequency.

The relevance of heat flux~\cite{Sekimoto2010} and entropy flux and production~\cite{Seifert2005} in characterizing OS is evident when we view OS as a cyclic irreversible thermodynamic process.	But, in the limit $\alpha \to 0$, these fluxes and production vanish, while the work done is a non-zero periodic function which vanishes only when time-averaged.

The house-keeping heat flux, at O($\alpha$), gets modified by the potential drive, and we find 
	\begin{align}
		q_\text{hk} = q_\text{hk}^{(0,0)} + k_1 q_\text{hk}^{(0,1)} + \cdots
		= \alpha \gamma (T_s - T_b) 
	\end{align}
where
	\begin{align}
	 q_\text{hk}^{(0,0)}=  \alpha \gamma ( \Sigma_{22}^{(0,0)}- T_b)~, \quad q_\text{hk}^{(0,1)} =  \alpha \gamma\Sigma_{22}^{(0,1)}~.
\end{align}
It is easy to see that the time-average of $q_\text{hk}$ continues to be zero in presence of the drive at $O(k_1)$, and hence there is no net energy exchange between the system and heat bath over a time period. In fact, the average heat-flux remains zero at $O(k_1)$ even if~$k(t)$ contains higher harmonics, and may become non-zero only when higher-order terms in$~k_1$ are considered. The properties of house-keeping entropy flux $\Phi_\text{hk} =q_\text{hk}/T_b$ can be read from $q_\text{hk}$.  

The entropy production rate~\cite{Seifert2005, Awasthi2022} is  given by the expression
	\begin{align}\label{ent-pr}
	\Pi :=\frac{1}{D_\alpha} \left\langle \left(\frac{\hat{J}_v^\text{ir}}{\hat{P}_\text{os}} \right)^2_t \right\rangle ~,
\end{align}
where the irreversible current
	\begin{equation}\label{irr-prob-curr}
	J_v^\text{ir} (x,v, t) = - \left( \gamma_\alpha v  + D_\alpha \frac {\partial}{\partial v} \right) P_\text{os}(x,v,t)~.
\end{equation} 
The house-keeping entropy production rate, required to sustain the OS, can be easily evaluated for Gaussian distributions $P_\text{os}(x,v,t)$ in terms its second moments. We find 
\begin{equation}
		\Pi_\text{hk} = \Pi_\text{hk}^{(1,0)} + k_1\Pi_\text{hk}^{(1,1)} +\cdots ~,
	\end{equation}
where 
\begin{equation}\label{en-prod-k1}
\Pi_\text{hk}^{(1,0)}=  \alpha D \left( \frac{\gamma}{D} - \frac{\overline{\gamma}}{\overline{D}} \right)^2 \Sigma_{22}^{(0,1)} ~, \quad	\Pi_\text{hk}^{(1,1)} = \alpha  D\left[ \left(\frac{\gamma}{D}\right)^2 - \left(\frac{\overline{\gamma}}{\overline{D}}\right)^2  \right] \Sigma_{22}^{(0,1)}~.
\end{equation}
	
Essentially, in the limit $\alpha \to 0$, there is no heat and entropy exchange between the system and the bath, while the potential drive pumps energy in and out of the system, maintaining it far from equilibrium with a time-periodic temperature. The interplay between entropy production and heat flux gets displayed only at $O(\alpha)$, where potential drives also participate actively in this background. 
	
	\section{Under the influence of time-dependent anharmonic perturbations}\label{chap7-sec3}
	We have described the characteristics of OS for driven Brownian particle in harmonic potential which is either time-dependent or has constant strength. These characteristics are not restricted to harmonic potentials alone. In this section, we investigate OS behavior in potential drives that include quartic perturbations. More specifically, we choose the $U(x,v,t) = k_0 x^2/2 + \lambda(t) x^4/4$, where the periodic function
	\begin{equation}
		\lambda(t) = \lambda_0 + \lambda_1 \cos(\omega t + \phi_\lambda)~,
	\end{equation} 
	with some phase difference$~\phi_\lambda$ from that of noise drive$~D(t)$, and evaluate thermodynamic properties in the low viscous regime to $O(\lambda)$.
	
	The OS distribution to $O(\lambda)$ takes the form
	\begin{equation}\label{P1}
		P_{os}(x,v,t) = \left[1 - \left( A^{(1)} - \langle A^{(1)} \rangle_0 \right)\right]P_{os}^{(0)}(x,v,t)~,
	\end{equation}
	where the $O(\lambda^0)$ term $~P_{os}^{(0)}$ is the harmonic OS distribution, the $O(\lambda)$ term 
	\begin{equation}\label{A1}
		A^{(1)} = \sum_{r=0}^2 \tilde{a}_r x^{2-r} v^r+ \sum_{r=0}^4 a_r x^{4-r} v^r~,
	\end{equation}
and $\langle A^{(1)} \rangle_0$ are average of $A^{(1)}$ with respect to$~P_{os}$. The $T$-periodic coefficients$~\tilde{a}_r$ and$~a_r$ satisfy certain dynamical equations which are established by substituting Eq.~\eqref{P1} in the corresponding Fokker-Planck equation of the anharmonically driven Langevin system, and then equating the coefficients of independent monomials to zero. These dynamical equations for the quartic potential reduce to 
	\begin{equation} \label{a4eom}
		\frac{d}{dt}	
		\begin{bmatrix}
			a_0\\ a_1\\ a_2\\a_3\\a_4
		\end{bmatrix}= 
		\begin{bmatrix}
			0 & k_p & 0 & 0 & 0\\
			-4 & \gamma_p & 2 k_p&0 & 0\\
			0 & -3 &2\gamma_p&3k_p & 0\\
			0 & 0 & -2 & 3\gamma_p & 4 k_p\\
			0 & 0 &0 &-1 & 4\gamma_p
		\end{bmatrix} 
		\begin{bmatrix}
			a_0\\ a_1\\ a_2\\a_3\\a_4
		\end{bmatrix}
		+ \lambda
		\begin{bmatrix}	
			\Sigma^{-1}_{12}\\ \Sigma^{-1}_{22}\\0\\0\\0
		\end{bmatrix},
	\end{equation}
	and
	\begin{equation} \label{a2eom}
		\frac{d}{dt}	
		\begin{bmatrix}
			\tilde{a}_0\\ \tilde{a}_1\\ \tilde{a}_2
		\end{bmatrix}= 
		\begin{bmatrix}
			0 & k_p & 0 \\
			-2 & \gamma_p & 2 k_p\\
			0 & -1 &2\gamma_p
		\end{bmatrix} 
		\begin{bmatrix}
			\tilde{a}_0\\ \tilde{a}_1\\ \tilde{a}_2	 
		\end{bmatrix}
		+ 2D
		\begin{bmatrix}
			a_2\\3a_3\\6a_4
		\end{bmatrix},
	\end{equation}
	where the $T$-periodic parameters
	\begin{align}\label{gp-kp}
		\gamma_p := \gamma - 2 D (\Sigma^{-1})_{22}~, \nonumber \\
		k_p := k - 2 D (\Sigma^{-1})_{12}~,
	\end{align}
	and $\Sigma_{ij}$ denote harmonic second-moments.

	Solving Eqs.\eqref{a4eom} and\eqref{a2eom} essentially amounts to solving fifth- and third-order ordinary differential equations, respectively. They can also be solved by following the method outlined in Ref.~\cite{Awasthi2021}, wherein the fundamental matrices of these equations are written in terms of solutions of a modified Hill equation 
	\begin{equation}\label{hill-eq-p}
		\frac{d^2}{dt^2} u_p + \nu_p u_p =0~, \quad \nu_p = k_p -  \frac{1}{2} \dot{\gamma}_p - \frac{1}{4} \gamma_p^2~.
	\end{equation}
We can employ the method to find the perturbative solutions $u_p=\sum_n \alpha^n u_p^{(n)}$, for the parameters 
	\begin{align}
		k_p(t) &= k_0 - 2 \alpha D(t) \Sigma_{\;12}^{-1}(t) = k_0 + 2 \alpha^2 D \frac{\Sigma_{12}^{(1)}}{\Sigma_{11}^{(0)}\Sigma_{22}^{(0)}} +\cdots~,\\
		\gamma_p(t) &= \alpha \gamma(t) - 2 \alpha D(t) \Sigma_{\;22}^{-1}(t)= \alpha \left(\gamma -  \frac{2  D}{\Sigma_{22}^{(0)}}  \right) + \alpha^2 \frac{\Sigma_{22}^{(1)}}{\left(\Sigma_{22}^{(0)}\right)^2}+\cdots~,
	\end{align}
	where $\Sigma_{12}^{(0)}$ and $\Sigma_{22}^{(0)}$ are the $O(\alpha^0)$ part of harmonic second moments. The parameter$~k_p$ differs from$~k$ only at $O(\alpha^2)$, while $\gamma_p$ is an $O(\alpha)$ quantity. Here too, periodicity at $O(\alpha)$ is required to establish the solutions at $O(\alpha^0)$. We find, which can also be verified directly substituting in Eq.~\eqref{a4eom}, the following leading-order expressions: 
	\begin{align}
		a_0^{(0)} &= \frac{\gamma_0 \lambda_0}{4 D_0}  - \frac{3 \gamma_0 D_1 \lambda_1 k_0 \cos (\phi_\lambda )}{16 D_0^2 \left(\omega ^2 - 4 k_0\right)} + \frac{\gamma_0 \gamma_1 k_0 \left(10 k_0-\omega ^2\right) \cos (\omega t +\phi_\lambda )}{D_0 \left(\omega ^2-16 k_0\right) \left(\omega ^2-4 k_0\right)}~,\\
		a_1^{(0)} &= \frac{\gamma_0 \gamma_1 \omega  \left(\omega ^2-10 k_0\right) \sin (\omega t +\phi_\lambda )}{D_0 \left(\omega ^2-16 k_0\right) \left(\omega ^2-4 k_0\right)}~,\\
		a_2^{(0)} &= \frac{3 \gamma_0 D_1 \gamma_1 \cos (\phi_\lambda )}{8 D_0^2 \left(4 k_0-\omega ^2\right)} + \frac{3 \gamma_0 \gamma_1 \cos (\omega t +\phi_\lambda )}{D_0 \left(\omega ^2-16 k_0\right)}~,\\
		a_3^{(0)} &= -\frac{6 \gamma_0 \gamma_1 \omega  \sin (\omega t +\phi_\lambda )}{D_0 \left(\omega ^2-16 k_0\right) \left(\omega ^2-4 k_0\right)}~,\\
		a_4^{(0)} &= -\frac{3 \gamma_0 D_1 \gamma_1 \cos (\phi_\lambda )}{16 D_0^2 k_0^2\left( \omega^2 - 4 k_0 \right)} -\frac{6 \gamma_0 \gamma_1 \cos (\omega t +\phi_\lambda )}{D_0\left(\omega ^2-16 k_0\right) \left(\omega ^2-4 k_0\right)}~.
	\end{align}
 These quantities appear in inhomogeneous part of Eq.~\eqref{a2eom}. The leading-order terms of arbitrary solution $\tilde{a}$ asymptotically become constants, since the $O(\alpha)$ inhomogeneous term drops out, and in fact turn out to be zero, $\tilde{a}_0^{(0)} =\tilde{a}_1^{(0)} = \tilde{a}_2^{(0)} = 0$, when periodicity is imposed at the next order. It should be noted that in order to calculate $O(\alpha^0)$ arbitrary solution $a_r^{(0)}$ of Eq.~\eqref{a4eom}, we need $O(\alpha^0)$ harmonic moments $\Sigma_{ij}^{(0)}$ which is fixed by demanding periodicity of first-order arbitrary solution $\Sigma_{ij}^{(1)}$. Hence in order to fix $a_r^{(0)}$, we need the arbitrary solution $a_r^{(1)}$ which in turn requires the corrections $\Sigma_{ij}^{(2)}$. Thus determining the first-order corrections $a_r^{(1)}$ for arbitrary drives is in general cumbersome. We list these corrections for a specific example.\\

	\noindent \textbf{An illustrative example for 1st order correction for anharmonic perturbation}\label{lambda2-expl}\\
	
	\begin{figure}[t]
		\includegraphics[width=0.6\linewidth]{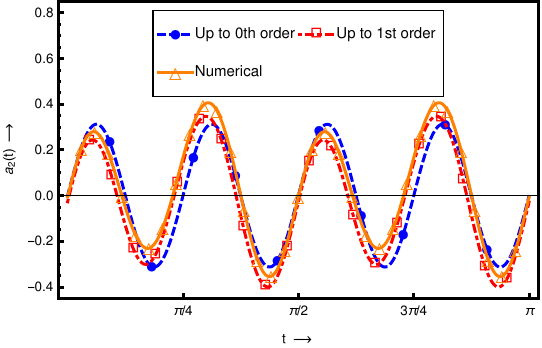}
		\caption{The asymptotic value of $a_{2} (t)$ has been plotted for two time periods. The viscous coefficient  $\gamma(t) = 5 + \cos(\omega t)$,  noise strength $D(t) = 3 + 2\sin(\omega t) + \sin(\omega t)$, drive frequency $\omega = 4$ , harmonic potential parameter $k_0 = 4$, anharmonic potential parameter $\lambda =2$  for $\alpha = 0.03$.}
		\label{smlalpAnh}
	\end{figure}
	Let us consider the example of $\gamma (t) = 5 + \cos(4 t), ~D(t) = 3 + 2 \sin(4 t), ~k_0=3,~$and$~\lambda = 2$. The $O(\lambda)$-corrections of the $a_r$-coefficients are found to be 
	\begin{align}
		(a_0)_1 =& -\frac{695}{576}-\frac{545}{128} \sin(4t)-\frac{2725}{192} \cos(4 t)\\
		(a_1)_1 =& \frac{1925}{144} \sin (4 t)-\frac{385}{96} \cos (4 t)\\
		(a_2)_1 =& -\frac{55}{96}-\frac{55}{64} \sin (4 t)-\frac{275}{96} \cos (4 t)\label{a2-1-smal-alp}\\
		(a_3)_1 =& \frac{275}{48} \sin (4 t)-\frac{55}{32} \cos (4 t)\\
		(a_4)_1 =& -\frac{55}{576}+\frac{55}{128} \sin (4 t)+\frac{275}{192} \cos (4 t)\\
		(\tilde{a}_0)_1 =&0~,\quad(\tilde{a}_1)_1 =0~,\quad(\tilde{a}_2)_1 =0
	\end{align}
	In Fig.~\ref{smlalpAnh}, we have plotted the asymptotic $a_2(t)$ for this example given in Eq.~\eqref{a2-1-smal-alp} and compared it to the numerical results that we obtain by directly solving the evolution equation. We notice that the analytical results have good agreement with the numerical results.\\
	
	\noindent \textbf{Thermodynamic properties of the oscillating state}\\
	
	Having obtained the OS distribution, we can now calculate various thermodynamic quantities of interest. The expectation of any observable$~g=g(x,v)$ in the OS can be expressed using Eq.~\eqref{P1} as
	\begin{align}\label{g-op-exp}
		\langle g\rangle
		=\langle g \rangle_0 - &\left[\tilde{a}_0 \langle g :x^2:\rangle_0
		+ \tilde{a}_1 \langle g :xv:\rangle_0 + \tilde{a}_2 \langle g :v^2:\rangle_0 + a_0  \langle g :x^4:\rangle_0\right. \nonumber \\
		& \left.  + a_1 \langle g:x^3 v:\rangle_0 + a_2  \langle g :x^2v^2:\rangle_0 +  a_3 \langle g:xv^3:\rangle_0 
		+ a_4 \langle g:v^4:\rangle_0 \right], 
	\end{align} 
	where$~:f:$ denotes$~{f-\langle f \rangle_0} $ for any function$~{f=f(x,v)}$.
	For small viscous drives, each term on the right-hand side of Eq.~\eqref{g-op-exp}  has to be Taylor expanded in $\alpha$. The explicit expression for the kinetic temperature, to $O(\alpha^0)$, is given by
	\begin{align}
		T_s(t) &= \langle v^2 \rangle_0 - 2 \tilde{a}_2 \langle v^2\rangle_0^2 - 12 a_4 \langle v^2 \rangle_0^3 - 2 a_2 \langle v^2 \rangle_0^2 \langle x^2\rangle_0\\
		& = \frac{D_0}{\gamma_0}\left[ 1 +\frac{3 \lambda_1}{k_0 \gamma_0 } \frac{ D_1 \cos (\phi_\lambda )-2 D_0 \cos (\omega t +\phi_\lambda )}{ \left(\omega ^2 - 4 k_0\right)}\right]~. \label{Ts-lt} 
	\end{align}
	Similarly, the configuration temperature with anharmonic corrections is obtained to be 
	\begin{align}
		T_c(t) &= k_0 \langle x^2 \rangle + 3 \lambda (t) \langle x^2 \rangle_0 \\
		& = k_0 \Sigma_{11}^{(0)}- 2 \tilde{a}_0^{(0)} k_0 (\Sigma_{11}^{(0)})^2 - 2 a_2^{(0)} k_0 \Sigma_{22}^{(0)} (\Sigma_{11}^{(0)})^2 - 12 a_0^{(0)} k_0 (\Sigma_{11}^{(0)})^3 \\
		& = \frac{D_0}{\gamma_0}\left[1 + \frac{3 \lambda_1}{\gamma_0 k_0^2}\left(\frac{ D_1 k_0 \cos (\phi_\lambda )}{\omega ^2 - 4 k_0}  +\frac{D_0 \left(\omega ^2 - 2 k_0\right) \cos (\omega t +\phi_\lambda )}{\omega ^2 - 4 k_0}\right)\right]\label{Tc-lt}~,
	\end{align}
which is distinct from kinetic temperature. In fact, the difference between the two temperatures 
	\begin{align}
		T_s -T_c = \frac{3 D_0^2 \lambda_1 \omega ^2  \cos (\omega t +\phi_\lambda)}{\gamma_0^2 k_0^2 \left(\omega ^2 - 4 k_0\right)}~,
	\end{align}
	is proportional to $\lambda_1$, which is induced by the potential drive as in the driven harmonic case.
	
	The correlation between position and velocity develops at the leading order due to the time-dependent $\lambda(t)$ and is given by 
	\begin{align}\label{c-lt}
		C_{xv} = -\frac{3 D_0 \lambda_1 \omega  \sin (\omega t +\phi_\lambda )}{\gamma_0 k_0^{3/2} \left(\omega ^2- 4 k_0\right)}~.
	\end{align}
	We notice, from Eqs.~(\ref{Ts-lt},~\ref{Tc-lt},~\ref{c-lt}), that the phase relationship between the correlation function, kinetic temperature, and configuration temperature is preserved even upon adding anharmonicity to the system. Essentially, in the limit $\alpha \to 0$, when all the drives $\lbrace \gamma(t),~D(t),~k(t),~\lambda(t)\rbrace$ are in phase with each other, then the correlation function is out of phase with them by $\pi/2$, unlike  the two temperatures which are in phase with them.
	
	At $O(\alpha)$, OS will require housekeeping heat and entropy flux to sustain. Time-dependent anharmonic potential is responsible for the work done, and at leading order its rate is given by
	\begin{align}
		\left\langle \frac{\dbar W}{dt} \right\rangle = -\frac{3 D_0^2 \lambda_1 \omega \sin (\omega t+\phi_\lambda )}{4 \gamma_0^2 k_0^2}~.
	\end{align}
 We also note that for given potential parameters, the rate of work done is higher when the unperturbed temperature $\overline{D}/\overline{\gamma}$ is higher. The entropy production rate remains zero at $O(\alpha^0)$, and at $O(\alpha^1)$ is given by 
	\begin{align}
		\Pi_\text{hk} = \Pi_\text{hk}^{(0,1)} - \frac{2 \alpha}{k_0}\left(\frac{D_0}{\gamma_0}\right)^3\left[\left( \frac{\gamma_0}{D_0}\right)^2 - \left( \frac{\gamma}{D}\right)^2 \right] a_\tau (t)~,
	\end{align}
	where $a_\tau(t) = a_2^{(0)} + 6 k_0 a_4^{(0)}$, and $\Pi_\text{hk}^{(0,1)} $ is given in Eq.~\eqref{en-prod-k1}. We notice that under anharmonic drive, similar to harmonic drive, the rate of entropy production becomes zero when both $D(t)$ and $\gamma(t)$ are independent of time. Essentially, potential drives alone will not produce entropy. This property though is true only to linear order in time-dependent potential. Furthermore, for time-independent anharmonic perturbation, namely when $\lambda_1=0$, we see that OS approaches equilibrium as $\alpha \to 0$.
		
	\bibliography{RefShakul4_PRL}